\documentclass[trackchanges,twocolumn]{aastex701}
\usepackage{amsmath}
\usepackage{pifont}

\usepackage{graphicx}	
\usepackage{amsmath}	
\usepackage{gensymb} 

\usepackage{xcolor}    
\usepackage[normalem]{ulem}  

\begin{document}

\newcommand{\fermi}{{\it Fermi}-LAT}
\newcommand{\gray}{$\gamma$-ray}
\newcommand{\grays}{$\gamma$-rays}

\title{Impact of Spectral Coverage on Parameter recovery in Blazar  Modeling}

\author[0000-0003-2011-2731]{N. Sahakyan}
\affiliation{ICRANet-Armenia, Marshall Baghramian Avenue 24a, Yerevan 0019, Armenia}
\email[show]{narek.sahakyan@icranet.org }  

\author[0000-0003-4477-1846]{D. B\'egu\'e}
\affiliation{Department of Physics, Bar-Ilan University, Ramat-Gan 52900, Israel}
\email{begueda@biu.ac.il}

\author[0000-0002-2265-5003]{P. Giommi}
\affiliation{Center for Astrophysics and Space Science (CASS), New York University Abu Dhabi, PO Box 129188 Abu Dhabi, United Arab Emirates}
\affiliation{Associated to INAF, Osservatorio Astronomico di Brera, via Brera, 28, I-20121 Milano, Italy}
\affiliation{Institute for Advanced Study, Technische Universit{\"a}t M{\"u}nchen, Lichtenbergstrasse 2a, D-85748 Garching bei M\"unchen, Germany}
\email{giommipaolo@gmail.com }

\author[0000-0002-8852-7530]{H. Dereli-B\'egu\'e}
\affiliation{Department of Physics, Bar-Ilan University, Ramat-Gan 52900, Israel}
\email{husne.dereli-begue@biu.ac.il}

\author[0000-0001-8667-0889]{Asaf Pe'er}
\affiliation{Department of Physics, Bar-Ilan University, Ramat-Gan 52900, Israel}
\email{asaf.peer@biu.ac.il}

\begin{abstract}
Understanding the impact of spectral coverage on parameter recovery is critical for accurate interpretation
of blazar spectra. In this study, we examine how the data coverage influences the reliability of parameter
estimation within the one-zone synchrotron self-Compton (SSC) framework. Using OJ 287, TXS 0506+056, and Mrk 421 as
representative of the low-, intermediate- and high synchrotron peak classes (LSP, ISP and HSP),
respectively, we generate synthetic SEDs based on their best-fit models and perform 1,000 fits for each of
the 21 observational configurations per source type. Our analysis quantifies the coverage probability
for all model parameters, such has the magnetic field strength and the electron luminosity, and reveals
that different blazar subclasses exhibit distinct sensitivities to spectral gaps. For LSPs, a minimal dataset
comprising optical/UV, X-ray, and GeV $\gamma$-ray bands is sufficient for robust parameter inference. In
contrast, ISPs and HSPs require broader spectral coverage to constrain the physical parameters. For ISP,
we find that reliable parameter recovery can be achieved with two different minimal band combinations: 
\textit{(i)} X-ray, high energy $\gamma$-ray, and very high energy $\gamma$-ray data, or \textit{(ii)} optical/UV,
X-ray, and high energy $\gamma$-ray data. For HSPs, the minimal configuration enabling reliable parameter
recovery includes the optical/UV, X-ray, and very high energy $\gamma$-ray bands. We discuss the role
of very high energy $\gamma$-ray observations, showing that they significantly enhance parameter recovery
for HSPs. Our results provide practical guidelines for designing optimized multi-wavelength observation
campaigns and for assessing the robustness of SSC model inferences under incomplete spectral coverage.

\end{abstract}

\keywords{\uat{Blazars}{164} --- \uat{BL Lacertae objects}{158} --- \uat{Astronomy data modeling}{1859} ----- \uat{High Energy astrophysics}{739} }

\section{Introduction} 

The study of blazars has entered a new era. The multiplication of multi-wavelength and multi-messenger
campaigns is drastically increasing the volume of available data, while the recent advances in both
machine learning and modeling enable detailed spectral analysis by fitting state-of-the-art numerical
models of blazar emission to observations \citep{BSD23, SBC24, TVP24}. These developments are expected
to lead to a deeper understanding of the physical processes occurring in this kind of
active galactic nuclei (AGN).

Blazars are a subclass of AGN characterized by relativistic jets that are closely
aligned with our line of sight \citep{1995PASP..107..803U}. Their emission spans the entire electromagnetic
spectrum, from radio wavelengths up to high-energy (HE; $>100$ MeV) and very high-energy (VHE;$>100$ GeV)
\grays\ \citep{2017A&ARv..25....2P}. One of the defining properties of blazars is their rapid variability,
which can occur on timescales as short as minutes \citep[e.g.,][]{2007ApJ...664L..71A, 2016ApJ...824L..20A, 2018ApJ...854L..26S,
2014Sci...346.1080A}, indicating that their emission
is produced in a compact region and that highly energetic processes are taking place in
their jets. Blazars are the dominant class of extragalactic sources in the \gray\ sky \citep{2022ApJS..260...53A},
making them key targets for current and upcoming multi-wavelength and multi-messenger observatories.

The multi-wavelength emission of blazars exhibits a characteristic double-peaked structure, with the low-energy
component peaking between the optical and X-ray bands, and the HE component extending above the X-ray band. The
origin of this emission is still under debate and is typically categorized based on the type of particles
responsible for the emission. In leptonic scenarios, where electrons are the primary radiating particles,
the low-energy peak is interpreted as synchrotron emission from relativistic electrons, while the HE component
is produced from inverse Compton scattering of synchrotron photons produced by the same electron population
\citep[synchrotron self-Compton, or SSC, ][]{GMT85, MGC92,BM96a}, or external photons originating from the
accretion disk, the broad-line region, or the dusty torus \cite[external Compton, or EC, ][]{DSM92,DS94, SBR94,
2000ApJ...545..107B}. Instead in hadronic models, the low-energy component is again explained by synchrotron
radiation from electrons, but the HE component arises from synchrotron emission by relativistic protons  \citep[][]{MP01}
or by secondary particles produced through proton interactions \citep[e.g. ][]{MB89, Man93, MPE03}. Recently, the
detection of VHE neutrinos spatially coincident with blazars \citep[e.g.,][]{2018Sci...361..147I, 2018Sci...361.1378I,
2018MNRAS.480..192P, 2023MNRAS.519.1396S} has renewed interest in hadronic and lepto-hadronic models
\citep[see, e.g.,][]{2013ApJ...768...54B, 2018ApJ...863L..10A,2018ApJ...864...84K, 2018ApJ...865..124M,
2018MNRAS.480..192P, 2018ApJ...866..109S, 2019MNRAS.484.2067R,2019MNRAS.483L..12C, 2019A&A...622A.144S,
2019NatAs...3...88G, 2022MNRAS.509.2102G}, which can naturally account for both the electromagnetic and neutrino
emission. However, we only focus on SSC models in this paper.

Blazars are commonly classified based on the peak frequency of their synchrotron component \citep{PG95,2010ApJ...716...30A}.
Sources with an observed synchrotron peak frequency $\nu_{\rm syn} < 10^{14}$ Hz are categorized as low-synchrotron-peaked (LSP or LBL)
blazars. Those with $10^{14} \ \text{Hz} < \nu_{\rm syn} < 10^{15}$ Hz are classified as intermediate-synchrotron-peaked
(ISP or IBL) blazars, and sources with $\nu_{\rm syn} > 10^{15}$ Hz are classified as high-synchrotron-peaked (HSP or HBL)
blazars. Blazars are also grouped based on the strength of their optical emission lines. Sources with strong, broad emission
lines are classified as flat-spectrum radio quasars (FSRQs), while those with weak or absent emission lines are identified
as BL Lacertae (BL Lac) objects \citep{1995PASP..107..803U}.

From a theoretical perspective, significant progress has been made in developing numerical models that can satisfactorily
describe the physical processes occurring in blazar jets. While these models account for a broad range of relevant mechanisms,
their application to large samples of sources or extensive datasets has been limited due to the computational cost associated
with scanning the high-dimensional parameter space. Due to the non-linear dependence of the spectra on the parameters, a fit is essential, as  the model parameters and their uncertainties cannot be inferred separately. Recently, the application of neural networks (surrogate models) to the
interpretation of observed data especially in the case of blazars, have addressed this limitation by enabling fast and
efficient fitting of spectral or temporal emission from relativistic sources \citep{BSD23, SBC24,TVP24}. This method allows
for thousands of fits to be performed in a reasonable time, whereas previously only a few—if any—model realizations could
be explored. This is different from the traditional methods relying on presenting a single model that "visually" matches the
observations, without performing any statistical inference \citep[see however][]{FDB08, RPG24, STL24, NAO25}.
The ability to efficiently perform statistical analyses and retrieve model parameters from state-of-the-art numerical
simulations marks a paradigm shift in the study of blazars and other high-energy astrophysical sources. For instance,
using this technique, we analyzed more than 700 SEDs of OJ 287 and investigated how spectral variations correlate with
different sets of model parameters \citep{HSB25}.

On the other hand, alongside the advancement of theoretical models, the quality of observational data is rapidly improving.
The current volume of data already enables detailed multi-wavelength and multi-epoch analyses of blazar emission
\citep[e.g.,][]{BSD23, NAO25, HSB25}. However, in many cases, contemporaneous coverage across the full electromagnetic
spectrum is not available. This incomplete coverage introduces uncertainty regarding how reliably model parameters can
be determined when some spectral bands are missing. This issue is critical, as these parameters directly inform our
understanding of the physical conditions within blazar jets. Furthermore, the SEDs of LSP, ISP, and HSP blazars differ
significantly. As a result, the absence of data in a given energy band does not impact all classes equally—certain
bands are more informative for specific source types. Addressing this issue is the main objective of this paper.
Using one representative blazar from each subclass (LSP, ISP, and HSP), we systematically investigate how the exclusion
of data from specific bands affects the outcome of spectral fitting. This analysis allows to quantify the role of
different energy bands in determining model parameters and provides guidance for evaluating the reliability of model
outputs based on the available data coverage.

The current study is focused on the issue of missing bands in the spectral
coverage of LSP, ISP, and HSP,  which within each group share common spectral properties outside of flaring states. In principle, missing spectral bands can also affect population-level studies
of LSP, ISP, and HSP but this is outside the scope of the present paper.

The paper is organized as follows. In Section \ref{sec:style}, we introduce the three blazars selected for this
study: OJ 287, TXS 0506+056, and Mrk 421. Section \ref{sec:SSC_model} describes the SSC model adopted for the analysis and
presents the modeling results for each source. In Section \ref{sec:synthetic_data}, we explain the procedure used to generate
synthetic datasets and how different combinations of spectral coverage are constructed to mimic various observational scenarios.
Section \ref{sec:results} presents the main results of our parameter recovery analysis. The discussion and conclusions are provided
in the Sections \ref{sec:dis} and \ref{sec:conc}, respectively.

\section{Source Selection} \label{sec:style}

The classification of blazars based on the location of the synchrotron peak—into LSPs, ISPs, and HSPs is not only a
phenomenological distinction but also reflects fundamental differences in the underlying physical processes. Each
subclass tends to show a typical range for both the synchrotron and inverse Compton peak frequencies and these
characteristic spectral properties provide valuable insights into the physical conditions within the jet, such as the
efficiency of particle acceleration, the energy distribution of emitting particles, the magnetic field strength, and the emission region size. 
Therefore, when modeling the SEDs of blazars belonging to different categories, it is
crucial to identify which model parameters can be reliably constrained and which remain poorly determined. This depends
largely on the availability and quality of observational data across specific energy bands. Since each blazar subclass
tends to exhibit emission features in distinct parts of the electromagnetic spectrum, the absence of data in key bands
can significantly affect the robustness of parameter estimation, particularly for those related to particle energy
distributions and magnetic field strength.

To explore the impact of data availability and the consequences of missing observational coverage on the determination
of model parameters, we selected representative blazars from each synchrotron peak category-LSPs, ISPs, and HSPs—and
modeled their SEDs. While individual sources may exhibit distinct emission properties, we note that, on average, their
spectral characteristics follow the general trends associated with their respective classes. The use of representative
sources from each category is justified as the aim of this study is not to determine the physical origin of the emission
in each blazar subclass, but rather to investigate how the presence or absence of observational data in specific energy
bands affects the ability to constrain key model parameters.

The Fourth Catalog of Active Galactic Nuclei detected by the \fermi{} \citep[Data Release 3;][]{2022ApJS..263...24A}
includes a total of 1,379 BL Lac-type blazars, among which 353 are classified as LSPs, 347 as ISPs, and 425 as HSPs.
The remaining 254 sources lack a reliable classification. Among the LSPs, we selected OJ 287, a bright blazar located
at a redshift of $z = 0.307$. Although OJ 287 exhibits a mixed behavior in the X-ray band—with the emission sometimes
corresponding to the tail of the synchrotron component—we selected a period that clearly shows LSP characteristics.
For the ISP category, we choose TXS 0506+056, at redshift $z = 0.337$, which was associated with a VHE neutrino detection during
a flaring episode and has since been continuously monitored across multi-walength bands. We note that TXS 0506+056 has been classified as a “masquerading BL Lac”.  However, the optical spectrum of  this source only shows extremely weak emission lines \citep{2019MNRAS.484L.104P}, implying that any external photon field must be faint compared to the bright Doppler-boosted jet, and does not significantly affect the SED modeling. For the HSPs, we selected
Mrk 421, at redshift $z = 0.031$, one of the most extensively studied blazars in the context of multi-wavelength
observations. These three source were selected due to their extensive multiwavelength coverage. However, any other source from these categories with similarly strong data coverage could have been used. We note that a small redshift is an asset for the analysis. Indeed, considering sources at larger distances would result in larger uncertainties in the data used for the fit, and even possibly their non-availability.

The data for OJ 287 and TXS 0506+056 were obtained from the Markarian Multiwavelength Data Center\footnote{\url{http://mmdc.am}}
\citep[\texttt{MMDC};][]{2024AJ....168..289S}, a novel database, developed in cooperation with the Firmamento
platform\footnote{\url{http://firmamento.nyuad.nyu.edu}} \citep[][]{2024AJ....167..116T}, that provides time-resolved
SEDs of blazars. The multiwavelength light curves of both sources were examined to identify epochs without strong flaring activity in the optical/UV, X-ray, and \gray\ bands. While mild variability is always present, these epochs correspond to comparatively low-activity states, free of pronounced flux enhancements that could modify the emission level and spectral shape. Selecting such periods reduces the risk of biasing the modeling results due to transient flares.
For Mrk 421, we used data from the multi-wavelength campaign in 2009, which provides extensive
coverage across the electromagnetic spectrum \citep{2011ApJ...736..131A}.

\begin{figure*}[ht!]
\centering
\includegraphics[width=0.98\textwidth]{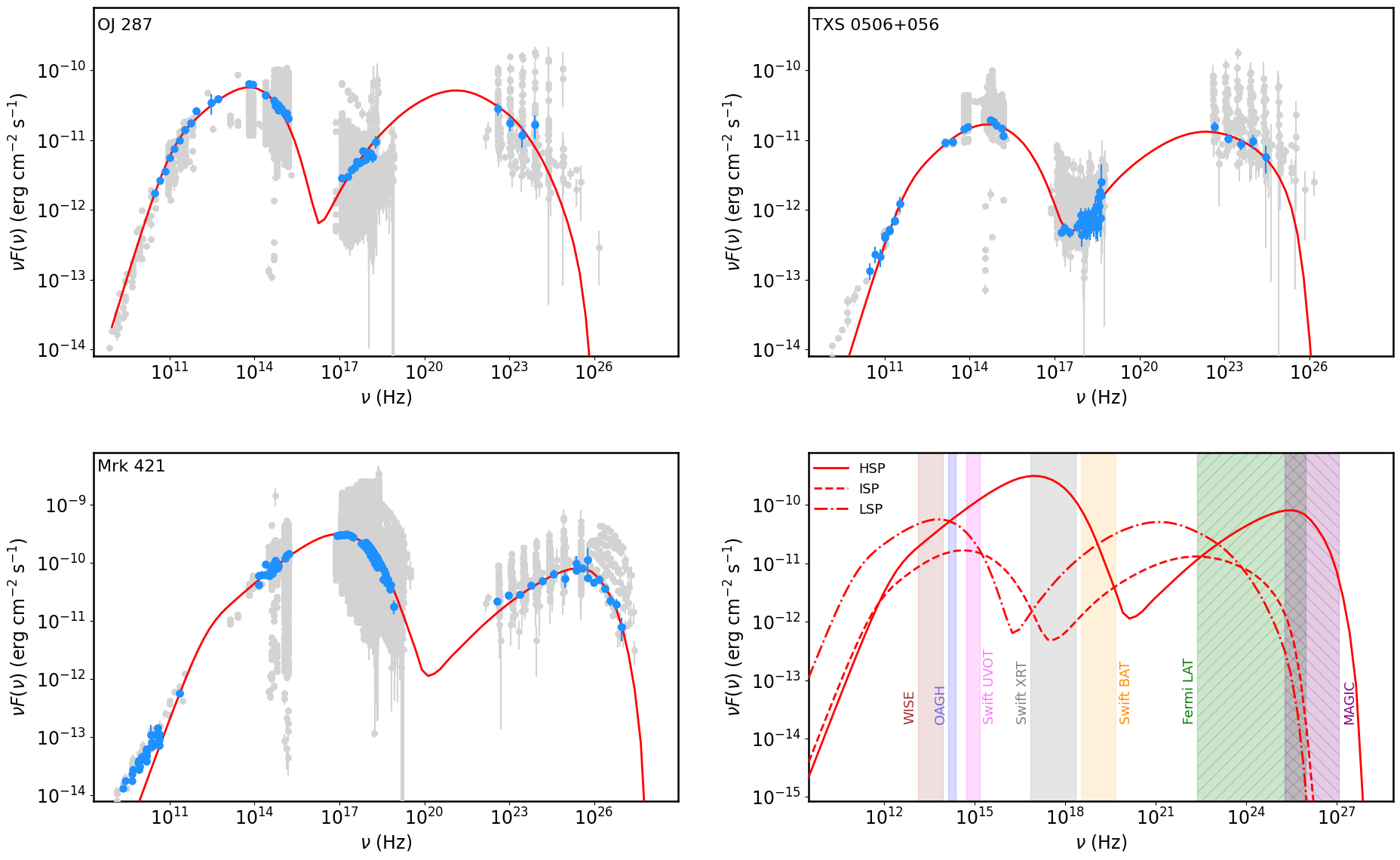}

\caption{Observed SEDs of the three selected sources. Top left: the LSP blazar OJ 287. Top right: the ISP blazar TXS 0506+056. Bottom
left: the HSP blazar Mrk 421. The data used for fitting are shown in blue, and the best-fit model is represented by the red
line. Archival data retrieved from \texttt{MMDC} and \texttt{Firmamento} are shown in gray. Bottom right: best-fit models for
all three blazars, displayed together with the instrumental bands used in this analysis. We note that the redshifts of these
blazars differ, and therefore their observed fluxes also strongly differ. Studying the systematics induced by the
different flux levels is beyond the scope of this work.}
\label{fig:sed_alongside_fit_results}
\end{figure*}

The broadband SEDs of OJ 287, TXS 0506+056, and Mrk 421 are shown in Figure \ref{fig:sed_alongside_fit_results}. The gray
points represent archival data extracted from MMDC/Firmamento databases, while the blue points correspond to the dataset
selected for modeling in the present study. As illustrated, the selected data represent an average emission state for each
source, reducing the influence of flaring episodes and providing a consistent picture of their emissions.

\section{The SSC model}
\label{sec:SSC_model}

The SED of blazars typically exhibits two broad components. In the case of BL Lacs, they are conventionally explained
using the SSC model \citep{GMT85, MGC92,BM96a} which is adopted in our study. Within this model, the lower-energy peak
is attributed to synchrotron radiation from relativistic electrons, whereas the HE peak arises from inverse Compton
scattering of these synchrotron photons by the same electron population. In this work we restrict our analysis to BL Lac objects; FSRQs, which require external Compton components, are not considered. We employ a one-zone SSC scenario in which all radiation is produced within a single homogeneous emitting region. This contrasts with two-zone or multi-zone models, where spatially distinct regions—typically associated with separate acceleration and cooling zones—can each contribute to the observed broadband emission.

In this study, we analyze the broadband emission from BL Lac type blazars using an advanced modeling technique
introduced by \citet{BSD23} which is available through the MMDC platform \citep{2024AJ....168..289S}. Unlike traditional
numerical computations, this method replaces computationally intensive numerical modeling with
a convolutional neural network (CNN) trained on radiative outputs of physical simulations. The CNN acts as a surrogate for the traditional numerical solver: it reproduces the same radiative solutions, preserves the underlying physics of particle injection and cooling, and has been validated against a large grid of numerical models \citep{BSD23}. Thus, the method is fully consistent with standard SSC modeling, differing only in computational efficiency. The methodology and
model setup are described in detail in \citet{BSD23}. The analogous external Compton modeling is described in \citet{SBC24}
and the lepto-hadronic model in \citet{2025ApJ...990..222S}. Below we briefly summarize the main assumptions of the model.

Within an SSC model, the emission is produced from a spherical emission region with radius $R$ and filled 
with a tangled magnetic field of strength $B$. Electrons are injected into this region at a rate described by a power-law spectrum
with an exponential cut-off:
\begin{align}
    \dot Q = Q_{e,0} \gamma^{-p} \exp \left( -\frac{\gamma}{\gamma_{\rm cut}} \right)
\end{align}
where $Q_{e,0}$ normalizes the injection rate, $p$ is the power-law index, $\gamma_{\rm min}$ is the minimum Lorentz
factor of the injected electrons, and $\gamma_{\rm cut}$ is the maximum Lorentz factor. These electrons cool
as they emit synchrotron emission and interact via these synchrotron photons by inverse Compton scattering.
The region is assumed to move relativistically toward the observer with a Lorentz factor $\Gamma$, and the jet is
viewed at an angle approximately $1/\Gamma$, leading to a Doppler amplification with a factor $\delta \approx \Gamma$. 

The model depends on seven free parameters: the electron spectral index $p$, the minimum and maximum Lorentz factors
of the injected electrons ($\gamma_{\rm min}$ and $\gamma_{\rm cut}$), the magnetic field strength $B$, the size of
the emitting region $R$, the Doppler factor $\delta$, and the electron luminosity $L_{\rm e}$, which is defined as:
\begin{align}
L_e = \pi R^2 \delta^2 m_e c^3 \int_1^{\infty} \gamma \, Q(\gamma) \, d\gamma.
\end{align}
where $Q$ is such that $\dot Q = Q/t_{\rm dyn}$, with $t_{\rm dyn} = R/ c $.
Note that the density of emitting electrons can be retrieved from the electron luminosity, Doppler boost and radius of the emitting region. The number of free parameters can be reduced when variability time is known but we keep the discussion general and do not restrict the analyses to pre-defined variability time scale. Instead we test how the spectral information only can constrain the parameters.

During the modeling, the first step is to retrieve the model parameters by optimization through maximization of the Gaussian likelihood function.
To explore the parameter space and obtain their posterior
distributions, we employ the MultiNest algorithm \citep{FHB09}, a nested sampling technique suitable for efficient Bayesian
inference in high-dimensional spaces. We use 1000 active points and set the sampling tolerance to 0.5 to ensure a balance
between computational efficiency and convergence accuracy.

The results of the SED modeling are shown in Figure \ref{fig:sed_alongside_fit_results}, with the model corresponding to the maximum likelihood (best-fit parameters) highlighted in red. The corresponding parameters
are provided in Table \ref{tab:param}. The SED of OJ 287, shown in the top left upper panel in Figure
\ref{fig:sed_alongside_fit_results}, is satisfactory modeled within the SSC model. The particle energy
distribution is characterized by a spectral index of $p = 2.18 \pm 0.2$, and the electrons are accelerated
up to a maximum Lorentz factor of $\log_{10}(\gamma_{\rm cut}) = 3.98 \pm 0.3$, with a minimum Lorentz factor of
$\log_{10}(\gamma_{\rm min}) = 2.37 \pm 0.3$. The magnetic field in the emitting region is
$\log_{10}(B/\mathrm{G}) = -1.71 \pm 0.2$, while the Doppler factor $\delta$ is estimated to be $\delta = 25.68 \pm 5.5$.
The radius of the emitting zone is inferred to be $\log_{10}(R/\mathrm{cm}) = 17.66 \pm 0.1$. The electron luminosity, 
$L_e = 1.35\times 10^{46}$ erg.s$^{-1}$ dominates over the magnetic luminosity
$L_{\rm B} = \pi c R^2 \delta^2 B^2 / 8\pi $ $\approx 2\times 10^{44}$ erg.s$^{-1}$, so the jet is particle-dominated
which is consistent with expectations for luminous BL Lac-type blazars.

In the case of TXS 0506+056 (upper right panel in Figure \ref{fig:sed_alongside_fit_results}), the SSC modeling similarly
provides an adequate modeling of the selected SED. The electron distribution exhibits a spectral slope of $p = 2.20 \pm 0.2$
and a minimum and  maximum Lorentz factor respectively of $\log_{10}(\gamma_{\rm min}) = 2.39 \pm 0.3$ and
$\log_{10}(\gamma_{\rm cut}) = 4.53 \pm 0.2$, indicative of highly energetic leptons within the jet. The magnetic field
strength is relatively low, with $\log_{10}(B/\mathrm{G}) = -1.01 \pm 0.3$, while the size of the emission region
is estimated to be $\log_{10}(R/\mathrm{cm}) = 16.98 \pm 0.3$. The Doppler factor derived from the modeling is
$\delta = 19.97 \pm 6.1$. The inferred luminosities again show a particle-dominated jet composition.

For Mrk 421, shown in the lower panel in Figure \ref{fig:sed_alongside_fit_results}, the observations reflects a
typical HSP blazar with strong X-ray and TeV components. The electron spectrum has an index of $p = 2.15 \pm 0.1$, with the
maximum electron Lorentz factor reaching $\log_{10}(\gamma_{\rm cut}) = 5.49 \pm 0.1$. The SSC fit constrains the magnetic
field to $\log_{10}(B/\mathrm{G}) = -1.32 \pm 0.2$, and the minimum electron Lorentz factor is found to be
$\log_{10}(\gamma_{\rm min}) = 2.68 \pm 0.2$. The emission region has a characteristic size of
$\log_{10}(R/\mathrm{cm}) = 16.44 \pm 0.3$, moving with a Doppler factor of $\delta = 34.04 \pm 5.6$. 

The peak frequency of the synchrotron component is set by the relation between
the magnetic field strength $B$ and the maximum electron Lorentz factor $\gamma_{\rm cut}$:
$\nu_p \propto \gamma_{\rm cut}^2 B$. From the results of our fits provided in Table
\ref{tab:param}, we find that $B$ remains steady between $10^{-2}$ and $10^{-1}$G with
rather large uncertainties, while $\gamma_{\rm cut}$ increases from $10^4$ to $10^{5.5}$
from the LSP to the HSP. This seems to indicate that, at least for the sources considered here,
the increase in observed frequency is primarily due to an increasing $\gamma_{\rm cut}$,
indicating more favorable particle acceleration properties for HSPs. A firm statistical
analysis of this interpretation is outside of the scope of the paper and will be investigated
elsewhere.

\begin{table*}
\caption{Model parameters describing the three selected SEDs shown in Figure \ref{fig:sed_alongside_fit_results}.}
\label{tab:param}
    \begin{tabular}{c||c||c|c}
                &  OJ 287 &   TXS 0506+056  & Mrk 421 \\\hline 
    Parameters  &  LSP    &   ISP           & HSP     \\\hline 
    \( p \)     & \( 2.18 \pm 0.2 \) & \( 2.20 \pm 0.2 \) & \( 2.15 \pm 0.1 \) \\
    \( \log_{10}(\gamma_{\rm cut}) \) & \( 3.98 \pm 0.3 \) & \( 4.53 \pm 0.2 \) & \( 5.49 \pm 0.1 \) \\
    \( \log_{10}(\gamma_{\rm min}) \) & \( 2.37 \pm 0.3 \) & \( 2.39 \pm 0.3 \) & \( 2.68 \pm 0.2 \) \\
    \( \delta \) & \( 25.68 \pm 5.5 \) & \( 19.97 \pm 6.1 \) & \( 34.04 \pm 5.6 \) \\
    \( \log_{10}(B/[\rm G]) \) & \( -1.71 \pm 0.2 \) & \( -1.01 \pm 0.3 \) & \( -1.32 \pm 0.2 \) \\
    \( \log_{10}(R/[\rm cm]) \) & \( 17.66 \pm 0.1 \) & \( 16.98 \pm 0.3 \) & \( 16.44 \pm 0.3 \) \\
    ~~~\( \log_{10}(L_{\rm e}/[\rm erg\:s^{-1}]) \)~~~ & ~~~ \( 46.13 \pm 0.2 \) ~~~ & ~~~ \( 45.00 \pm 0.3 \) ~~~ & ~~~ \( 43.77 \pm 0.1 \) ~~~ \\\hline
    \(\log_{10}(L_{\rm B}/[\rm erg\:s^{-1}]) \) & \( 44.30 \) & \( 44.11 \) & \( 42.88 \) \\
  \end{tabular}
\end{table*}

\section{Synthetic data Generation and fit}
\label{sec:synthetic_data}

The modeling results shown in Figure \ref{fig:sed_alongside_fit_results}, along with the corresponding parameters
listed in Table \ref{tab:param}, demonstrate that the different subclasses of blazars—LSP, ISP, and HSP—are
characterized by distinct physical conditions and electron populations responsible for the emission. However, we
do not perform a direct comparison of the physical parameters across different blazar types, as the goal of this work
is not to interpret the intrinsic physical differences between these sources. Instead, our focus is on understanding
how the availability or absence of observational data in specific energy bands influences the reliability of parameter
estimation. The parameters listed in Table \ref{tab:param} are used as reference values, and for each blazar type,
synthetic SEDs are generated and modeled using the same theoretical framework. A statistical analysis is then performed
to evaluate which parameters can be reliably constrained and which remain poorly determined under different observational
conditions.

\begin{table*}[t]
\caption{Definitions of observation bands in each of the 21 cases used in the analysis performed in this paper.}
\label{tab:obs_bands}
\begin{tabular}{c|c|c|c|c|c|c|c}
\hline
Case & WISE & OAGH & Swift-UVOT & Swift-XRT & Swift-BAT & \fermi{} & MAGIC \\
\hline
1  & No  & No  & No  & Yes & No  & Yes & Yes \\ \hline
2  & No  & No  & Yes & No  & No  & Yes & Yes \\ \hline
3  & No  & No  & Yes & Yes & No  & No  & Yes \\ \hline
4  & No  & No  & Yes & Yes & No  & Yes & No  \\ \hline
5  & Yes & No  & No  & Yes & No  & Yes & Yes \\ \hline
6  & Yes & No  & Yes & No  & No  & Yes & Yes \\ \hline
7  & Yes & No  & Yes & Yes & No  & No  & Yes \\ \hline
8  & Yes & No  & Yes & Yes & No  & Yes & No  \\ \hline
9  & No  & No  & Yes & Yes & No  & Yes & Yes \\ \hline
10 & No  & No  & Yes & Yes & Yes & Yes & No  \\ \hline
11 & No  & Yes & Yes & Yes & No  & Yes & No  \\ \hline
12 & Yes & No  & Yes & Yes & Yes & Yes & No  \\ \hline
13 & Yes & Yes & Yes & Yes & No  & Yes & No  \\ \hline
14 & No  & No  & Yes & Yes & Yes & Yes & Yes \\ \hline
15 & No  & Yes & Yes & Yes & No  & Yes & Yes \\ \hline
16 & No  & Yes & Yes & Yes & Yes & Yes & No  \\ \hline
17 & Yes & Yes & Yes & Yes & No  & Yes & Yes \\ \hline
18 & Yes & Yes & Yes & Yes & Yes & Yes & No  \\ \hline
19 & Yes & No  & Yes & Yes & Yes & Yes & Yes \\ \hline
20 & No  & Yes & Yes & Yes & Yes & Yes & Yes \\ \hline
21 & Yes & Yes & Yes & Yes & Yes & Yes & Yes \\
\hline
\end{tabular}
\end{table*}

In order to investigate the role of specific energy bands in parameter estimation, we selected the following instruments: 
\begin{itemize}
\item Wide-field Infrared Survey Explorer (WISE) - Infrared photometry data were used across four bands (3.4, 4.6, 12, and 22 $\mu$m) \citep{2010AJ....140.1868W}.
\item Observatorio Astrofísico Guillermo Haro (OAGH) - Near-infrared observations in the $H, J, K$ bands \footnote{http://astro.inaoep.mx/en/observatories/oagh/}.
\item Ultraviolet/Optical Telescope (UVOT) onboard the Neil Gehrels Swift Observatory (hereafter Swift) - Ultraviolet measurements in UVW1, UVM2, and UVW2 filters, and optical measurements in the $V, B, U$ bands \citep{2005SSRv..120...95R}.
\item X-ray Telescope (XRT) onboard Swift - X-ray observations in the 0.3–10 keV range \citep{2005SSRv..120..165B}.
\item Burst Alert Telescope (BAT) onboard Swift - Hard X-ray observations covering the 14–195 keV range \citep{2013ApJS..209...14K}.
\item Large Area Telescope (LAT) onboard the Fermi Gamma-ray Space Telescope (\fermi) - \gray{} data in the 0.1–400 GeV range \citep{2009ApJ...697.1071A}.
\item Major Atmospheric Gamma Imaging Cherenkov Telescopes (MAGIC) - VHE \gray{} observations in the 0.08–5.0 TeV range \citep{2016APh....72...76A}.
\end{itemize}
These bands were chosen as they provide coverage across a broad range of wavelengths, as can be seen in the bottom right
panel of Figure \ref{fig:sed_alongside_fit_results}.

It is important to note that this study is not focused on the specific instruments listed above, but rather on the energy
ranges they represent. Our aim is to assess the influence of data obtained in the infrared, optical, ultraviolet, X-ray, and \gray\ band 
on parameter estimation, independent of the specific characteristics of individual instruments. For instance, the conclusions
drawn here would remain valid if data from other instruments operating in similar bands were used—such as replacing Swift-UVOT
with any other ultraviolet or optical instrument, Swift-BAT with NuSTAR, or MAGIC with H.E.S.S., VERITAS or the upcoming
Cherenkov Telescope Array Observatory (CTAO). We emphasize that our analysis is performed on already processed
data in the $\nu$–$\nu F_\nu$ representation.
Therefore, instrument-specific observational methods, calibration procedures, or raw data systematics are not taken into account
in this analysis. If we performed the modeling on observed data (forward-folding), the choice of instrument and its technical
characteristics would indeed play a critical role. This is an important aspect that should be tested in the future. By working
with homogenized spectral data products, we ensure that our conclusions reflect the impact of spectral coverage rather than
instrument-specific effects. 

The bottom right panel of Figure~\ref{fig:sed_alongside_fit_results} shows the energy ranges covered by the
selected instruments, overlaid with representative SED models for LSP, ISP, and HSP blazars, shown as dot-dashed, dashed,
and  solid lines, respectively. It is evident that the selected bands sample different regions of the SED depending on
the blazar type. For instance, infrared to optical/UV data trace the rising part of the synchrotron component in HSPs,
while the same bands encompass the synchrotron peak for ISPs and the HE tail of the synchrotron component in LSPs.
Similarly, X-ray observations probe the HE tail of the synchrotron emission in HSPs, the transition region between
synchrotron and inverse Compton components in ISPs, and the rising segment of the inverse Compton component in LSPs.
HE and VHE \gray{} data span the entire inverse Compton component in HSPs but only capture its high-energy tail in LSPs
and the peak of HE component in ISPs. As a result, the inclusion or exclusion of data from specific bands affects the parameter
estimation differently depending on the blazar subclass. This highlights the importance of spectral coverage tailored
to the source type when modeling broadband emission.

\subsection{Mock data generation}
\begin{figure*}[ht!]
\centering
\includegraphics[width=0.98\textwidth]{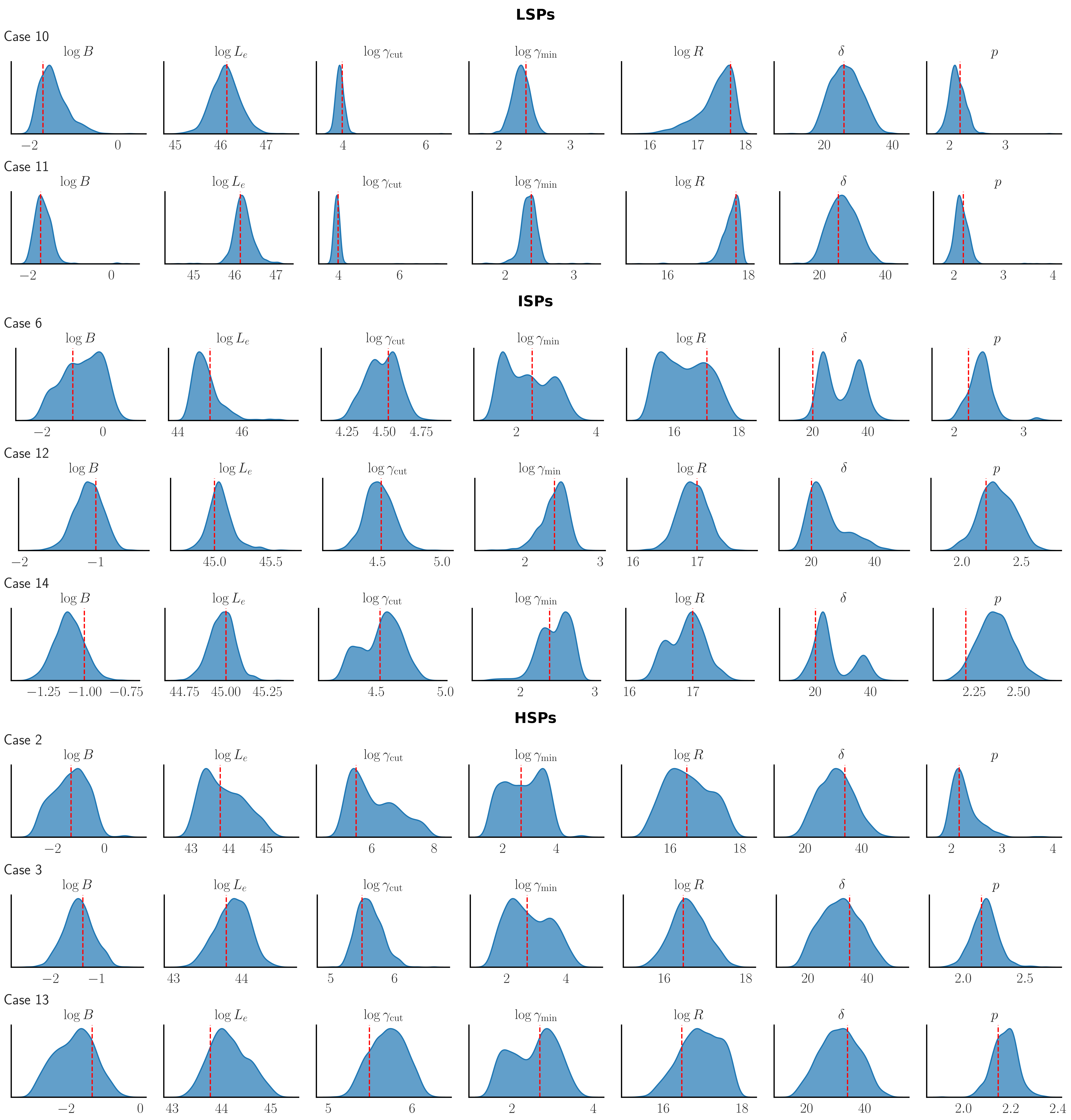}

\caption{ The density plots for selected LSP (top two rows), ISP (three middles rows), and HSP (three bottom rows) cases. The cases were chosen to show for each blazar type an acceptable recovery (HSP - case 3, ISP - case 12, LSP - case 11), a mediocre one (HSP - case 2, ISP - case 6) and an intermediate one (HSP - case 13, ISP - case 14, LSP - case 10) see the text for the details. The true parameter value, determined from the fits (see Table \ref{tab:param}), is shown as the red dashed line. If the distribution is narrowly peaked around the red dashed line at its center, the corresponding parameter is likely adequately recovered. Note that this is in fact quantified by the coverage probability.}
\label{fig:kde}
\end{figure*}

To assess the robustness of the modeling approach under varying observational constraints, we constructed a set of
representative cases encompassing a broad range of multi-wavelength data availability. These cases, summarized in
Table \ref{tab:obs_bands}, were designed to simulate different levels of data coverage, from sparse to dense. In this analysis, we assume that all data are quasi-simultaneous, similarly to the assumption we made to analyze the three dataset in Section \ref{sec:style}. In other words, spectral variation is not considered.
At the low-coverage end, cases such as Case 1 and Case 2 include only a minimal subset of bands, combining, for example,
X-ray and HE \gray{} observations (Swift-XRT, \fermi, and MAGIC), or ultraviolet and \gray{} data (Swift-UVOT, \fermi,
and MAGIC), respectively. These limited configurations are meant to reflect real-world scenarios where only partial simultaneous data
are available. Intermediate coverage scenarios are represented by cases such as Case 10 and Case 14, which include
contributions from most key instruments across infrared, optical, X-ray, and \gray{} bands, but may lack coverage in one
or two specific domains (e.g., MAGIC or Swift-BAT). Finally, Case 21 provides an example of full multi-wavelength coverage,
incorporating data from all considered instruments — WISE, OAGH, Swift-UVOT, Swift-XRT, Swift-BAT, \fermi, and MAGIC — allowing
for the most comprehensive characterization of the source SED. This stratified sampling of data availability enables
evaluation of the performance and reliability of the analysis framework under conditions ranging from highly constrained
to optimally sampled observations. In principle, other configurations with available data can also be generated; however, we have selected scenarios ranging from minimal datasets to maximal coverage, that are most relevant (1) for the spectral analysis, and (2) with respect to available multi-wavelength data.

For each configuration in Table \ref{tab:obs_bands} synthetic SEDs were generated to reflect the instrument-specific
coverage. For each observational band, the synthetic flux values were derived by interpolating the underlying model
SED (obtained from initial fit to the blazar data) at the logarithmic frequencies representative
of the respective instruments. In the case of WISE, OAGH and UVOT the data were generated at exact frequencies covered
by these instruments whereas five logarithmically spaced points were used within the BAT (14–195 keV), \fermi\ (0.1–400 GeV),
and MAGIC (80–5000 GeV) bands. For the Swift-XRT band (0.3–10 keV), data were generated at logarithmically spaced
frequencies with a step size of 0.1 in log-frequency space.

Uncertainties were implemented by adding Gaussian-distributed scatter to the interpolated flux values, with the standard
deviation scaled to mimic realistic observational errors. In the X-ray and \gray{} domains, where background-dominated
noise increases at higher energies, the fractional error was modeled as a linear function in log-frequency space, ranging
from 5\% at the lowest energies to 30\% at the high end of each energy range. This energy-dependent uncertainty model was
defined analytically for each instrument (e.g., Swift-XRT, Swift-BAT, \fermi, MAGIC) via a simple
linear relation:
\begin{equation}
\sigma_{\log F} = a \cdot \log(\nu) + b,
\end{equation}
with the slope $a$ and intercept $b$ computed to match the desired error scaling. In contrast, the optical and UV
bands (OAGH and Swift-UVOT) were assigned a uniform uncertainty of 5\%, reflecting their typically lower signal-to-noise
ratios in standard exposures. This approach ensures that each synthetic dataset not only respects the spectral coverage
of the selected instruments but also reflects realistic observational noise characteristics. Consequently, it enables
a controlled yet physically grounded assessment of model performance across the full spectrum of data completeness.

\subsection{Data fitting}

For each case 1,000 synthetic SEDs were generated. The 21 cases of different band observations were then created from
these 1000 SEDs, corresponding to a total of 21,000 realizations per blazar subclass (LSP, ISP, and HSP). This means that
the same data realization is used in all cases, which only vary by which bands are being considered. The fitting procedure
was executed on a local server utilizing a Docker-based implementation of the modeling code, publicly available through the
MMDC platform\footnote{\url{www.mmdc.am}}. The system was configured to process approximately 3,000 fits per hour, enabling
completion of all 63,000 fits in less than 24 hours.

For illustration, we show on Figure \ref{fig:kde} the kernel density estimations (KDE) of all recovered parameters for a
selected few cases for each blazar type. The red dashed lines shows the true parameter values, determined from the fits (see Table \ref{tab:param}). If it lies close to the maximum of the KDE, or at the center of the distribution, it is likely that the corresponding parameter is adequately recovered. Note that this is quantified by the coverage probability.

\section{Results}
\label{sec:results}

To assess the impact of spectral coverage on the reliability of model parameter recovery, we analyzed the fit
results of the synthetic SEDs generated under various observational configurations. To quantify the accuracy of
parameter estimation, we use the coverage probability, defined as the fraction of the 1000 synthetic fits whose 1$\sigma$
error intervals (i.e., $\text{fit value} \pm \sigma_{\text{error}}$) contain the true parameter value. The coverage
property is a number between 0 and 1, where a value close to 1 (at least above the threshold of 0.68) represents a good coverage, meaning that most of the
1000 fits retrieved the expected values within 1$\sigma$, while a value close to 0 means that none of the fit recovered
the right parameter value within 1$\sigma$, implying either a too constrained parameter value for which the
error does not capture the generative value or an unreliable parameter estimation.

\subsection{Low synchrotron peak: OJ287}

\begin{figure*}[t]
\centering
\includegraphics[width=0.98\textwidth]{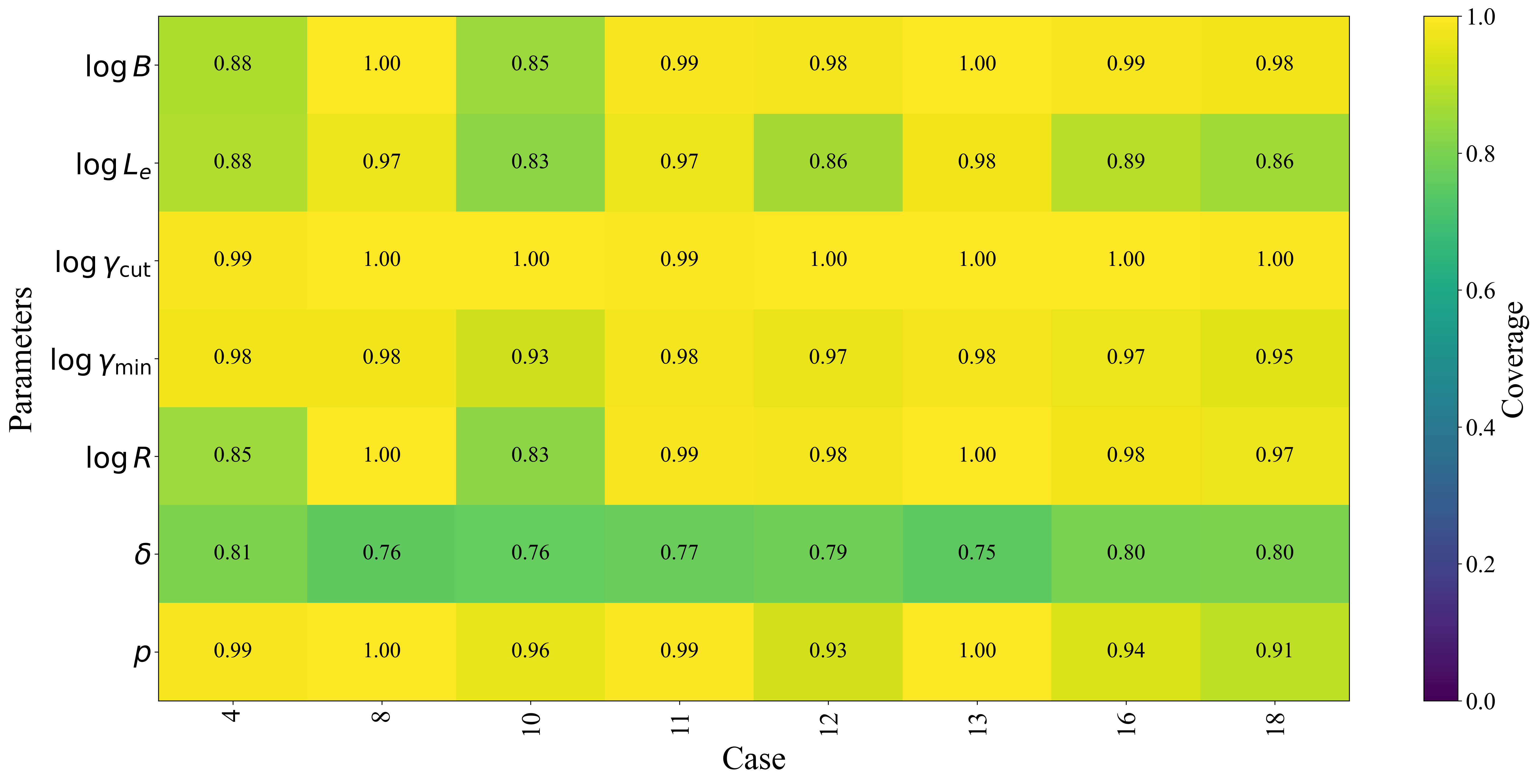}
\caption{Density matrix for the LSP OJ 287 showing the coverage probability for each parameter (line) in each case (column). The number in each cell corresponds to the coverage probability, which is also encoded in the colormap. Yellow cells correspond to high coverage probability while green cells represent lower coverage probability. A high coverage probability means that the fit retrieve the true parameter within its uncertainty. For LSPs, only cases without MAGIC data are presented.}
\label{fig:LSP_coverage}
\end{figure*}

For LSP sources, observations in the VHE $\gamma$-ray band are typically not available due to 
low flux levels, as the HE peak is in the MeV band. We therefore remove all cases considering
the MAGIC band for this class of blazar, leaving a total of 9 configurations for the LSP
scenario. The coverage probabilities for all model parameters across the remaining cases are
summarized in Figure \ref{fig:LSP_coverage}. Overall, the results indicate good coverage for almost
all parameters, with values typically exceeding 0.9 for most parameters. Only the Doppler factor
can be improperly retrieved, with a coverage probability dropping to $\sim 0.8$ across cases. 
This is explained by the fact that all considered cases includes X-ray (Swift-XRT) and $\gamma$-ray
(\fermi ) data together with low energy data, which are critical for constraining the HE component of the SED which in turn
determines parameters such as $\log \gamma_{\rm cut}$, $\log \gamma_{\rm min}$, and $p$, as well
as the normalization of the electron distribution. The latter, in combination with the magnetic
field strength, is further constrained by lower-energy data.

It is interesting that the coverage for $\log \gamma_{\rm cut}$ and $p$ reaches or approaches unity across all cases,
indicating a highly reliable recovery within 1$\sigma$ of these parameters under the SSC framework. In contrast,
the Doppler factor $\delta$ shows lower coverage values, ranging from 0.75 to 0.81, which shows a reduced sensitivity
of this parameter to the available data. This suggests that, with the current observational data in the 9 cases,
$\delta$ remains poorly constrained for LSP sources. As an illustrative example, the density distributions for Cases
10 and 11 is shown in the upper panel of Figure \ref{fig:kde}. The red dashed lines shows the
true parameter values, which lie near the centers of the corresponding density distributions in the most cases, confirming the high
coverage levels. 

\subsection{Intermediate synchrotron peak: TXS 0506+056}

\begin{figure*}[t]
\centering
\includegraphics[width=0.98\textwidth]{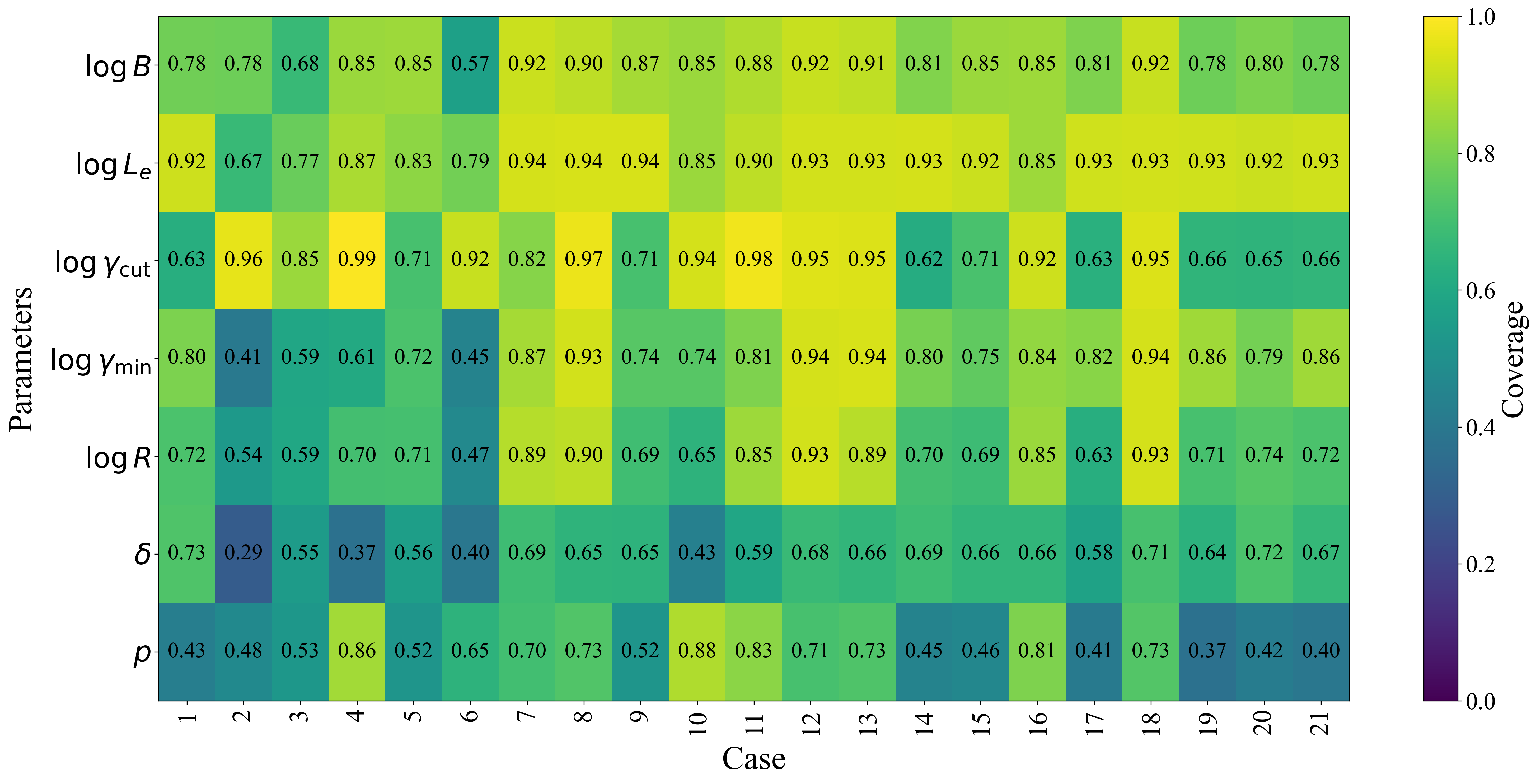}
\caption{ Same as Figure \ref{fig:LSP_coverage} but for the ISP TXS 0506+056. }
\label{fig:ISP_coverage}
\end{figure*}

For ISP sources, the synchrotron peak is observed at intermediate frequencies,
$10^{14} {\rm Hz} < \nu_{\rm syn} < 10^{15} \rm{Hz}$, and the inverse Compton
component can extend into the GeV–TeV band. Therefore, reliable parameter estimates for ISPs depend on capturing both
the synchrotron and inverse Compton components with sufficient spectral coverage. In contrast to LSPs, the
inclusion of VHE \gray-ray data is necessary for ISPs due to the higher flux levels expected in this band. Therefore,
all 21 observational configurations were analyzed, and Figure \ref{fig:ISP_coverage} shows the resulting parameter
coverage across all cases.
The strongest constraints are found in Cases 8, 12, and 18, where five out of seven model parameters reach coverage
above 0.9, as well as in Case 13, where four out of seven parameters exceed this threshold. These cases include
broad multi-wavelength coverage ranging from the infrared to TeV bands. Case 8 includes WISE, Swift-UVOT, SWIFT-XRT
and \fermi. Case 12 includes WISE, Swift-UVOT, Swift-XRT, Swift-BAT and \fermi. Case 13 includes WISE, OAGH,
Swift-UVOT, Swift-XRT, and \fermi\ data, while Case 17 includes a similar set with MAGIC data. This configuration
enables strong constraints on the parameters $\log \gamma_{\rm cut}$, $\log \gamma_{\rm min}$, $\log B$, $\log L_e$
and $\log R$, as both the low- and high-energy slopes of the SED are well constrained. It is interesting that coverage
for $\log R$ and $\log \gamma_{\rm min}$ also reach values above 0.9 in these cases, indicating that coverage across
multiple energy bands can actually help to constrain geometrical and low-energy cutoff parameters. Relatively strong
parameter recovery is also observed in Case 7, where all parameters exceed the 0.68 threshold, and $\log B$ and $\log L_e$
are constrained with coverage $\geq 0.9$.

Among all model parameters, $\log L_e$ and $\log \gamma_{\rm cut}$ are generally the best recovered, with coverage
$\geq 0.9$ in 14 and 10 cases, respectively. The parameter $\log B$ also shows strong performance, exceeding 0.9 in
five cases (Cases 7, 8, 12, 13, 18). In contrast, the Doppler factor $\delta$ and the electron spectral index $p$ are the
most difficult parameters to recover, with minimum coverage values of 0.40 and 0.37, respectively. For example low
coverage for $p$ is observed in Case 19, alongside with Case 20 and 21. This is interesting as these three cases have
the most comprehensive coverage of all our cases, only lacking a low-energy band for Case 19 and 20. For these three
cases, the fitted values of the spectral index is found to be systematically larger than the generative one, while the
error on this parameter is systematically small, resulting in a poor coverage probability. Since the data and their errors
are generated independently for each instrument in our mock observation, this shows that each band is important and should
be treated consistently to avoir systematic error in the fit.

Overall, the Doppler boost $\delta$ is poorly constrained in the majority of cases, with only 6 out of 21 exceeding the
0.68 coverage threshold. This indicates that flux-based SED fitting alone is insufficient to robustly determine the
Doppler factor and that additional information—such as variability timescales or source geometry—is required. The weakest
overall parameter recovery is observed in case 6, where only two parameters achieve coverage above 0.68 and the others
are below. Case 3 also performs relatively poorly, with no parameters reaching coverage above 0.9 and only three exceeding
the 0.68 threshold.

Figure \ref{fig:kde} (middle panel) presents the parameter distributions for Cases 6, 12, and 14. Case 12 demonstrates
well-peaked density distributions for most parameters while Case 6 shows broad distributions, with only $\log \gamma_{\rm cut}$
constrained. Case 14 represents an intermediate scenario: some parameters, such as $\log L_e$ and $\log B$, are well
constrained, while others remain weakly constrained due to gaps in observational coverage.

\subsection{High synchrotron peak: Mrk 421}

\begin{figure*}[ht!]
\centering
\includegraphics[width=0.98\textwidth]{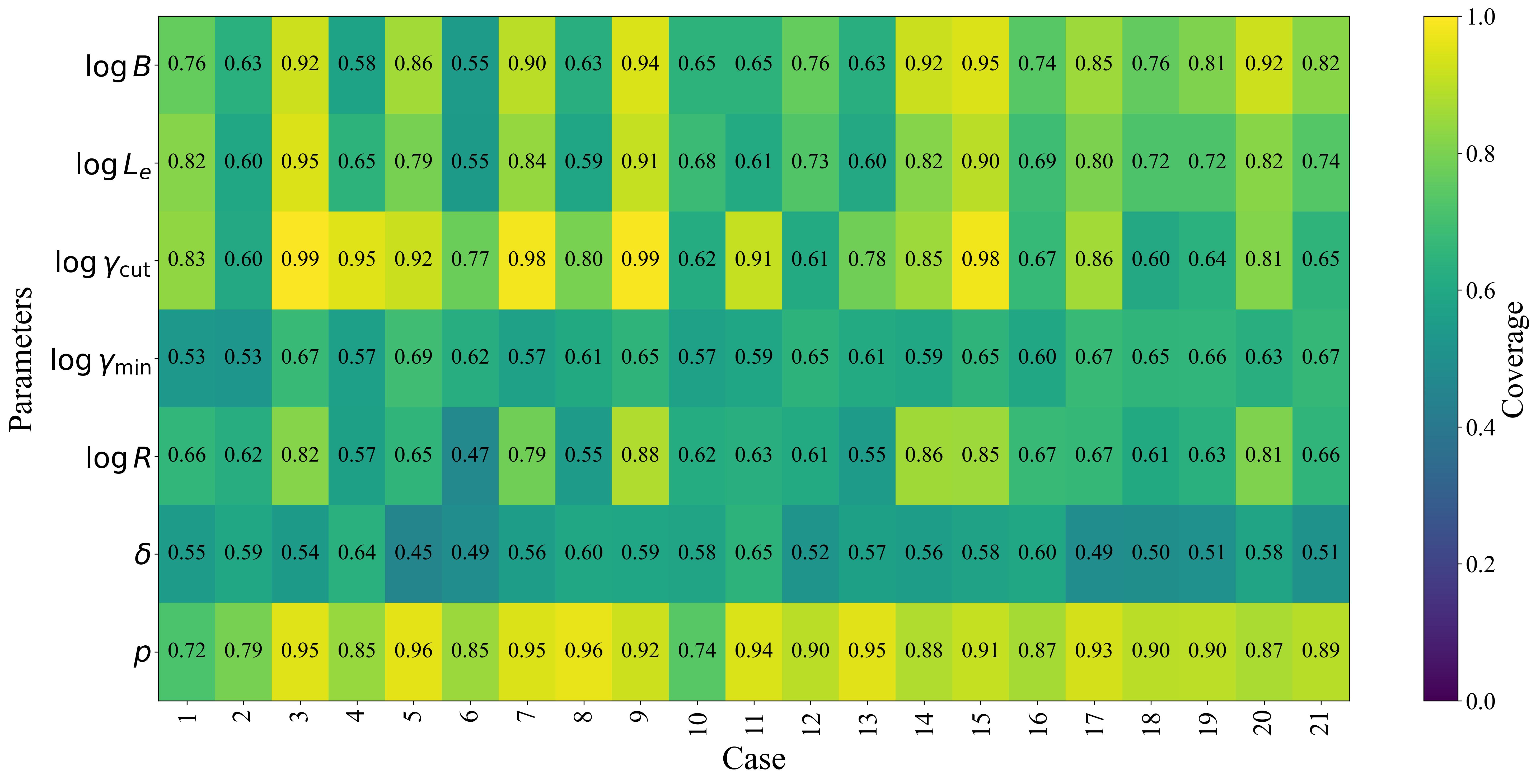}
\caption{Same as for Figures \ref{fig:LSP_coverage} and \ref{fig:ISP_coverage}, but for HSP. }
\label{fig:colormap_HSP}
\end{figure*}

For HSP sources, the synchrotron peak typically lies in the UV or X-ray band, resulting in the HE component extending from
a few GeV up to multi-TeV energies. Consequently, compared to LSP and ISP sources, observations in the X-ray and \gray\ bands,
especially in the VHE \gray\ band, are essential for performing reliable SED modeling and parameter estimation. All
21 observational configurations were included in the analysis, allowing for a systematic assessment of the contribution of
different data bases to the modeling and parameter recovery. Figure \ref{fig:colormap_HSP} presents the coverage probabilities
for all model parameters across the 21 considered cases, and the KDE of a few selected cases (2, 3, 13) are shown
in Figure \ref{fig:kde}.

The spectra of HSP sources have the broadest spectrum among the three source classes (LSP, ISP, HSP), leading to improved
recovery of several model parameters. The electron spectral index $p$ is the most reliably constrained parameter, with coverage
exceeding 0.9 in 12 out of 21 cases. In particular, very high coverage ($>0.95$) is achieved in Cases 3, 5, 7, 8 and 13.
These cases consistently include one band at low, intermediate, and high energy. Since the rising slopes of both the
synchrotron and inverse Compton components are sensitive to $p$, and since each of these components are covered by
the set of instruments, a robust parameter estimation is possible. The cutoff energy $\log \gamma_{\rm cut}$ is also
well retrieved in seven cases ($>0.95$), with coverage above 0.95 in Cases 3, 4, 7, 9, and 15. These configurations
include \fermi\ and/or MAGIC data at high energy, as well as Swift-UVOT and Swift-XRT, allowing constraints
on $p$ and $\log \gamma_{\rm cut}$ all-together. The LAT data primarily constrain $p$, while MAGIC or XRT data help
determine the cutoff energy. Another parameter with high coverage ($>0.9$) is the magnetic field strength $\log B$, which is
well constrained in Cases 3, 7, 9, 14, 15 and 20. These cases share a common observational setup, incorporating Swift-UVOT,
Swift-XRT and \fermi\ data. This combination enables simultaneous constraints on both $p$ and $\log \gamma_{\rm cut}$
which, in turn, defines the synchrotron and inverse Compton peaks allowing to estimate $\log B$.

In contrast, parameters such as $\log \gamma_{\rm min}$, $\log R$, and $\delta$ are poorly constrained. For
$\log \gamma_{\rm min}$, the coverage ranges from 0.53 to 0.69 in most cases, exceeding the 0.68 threshold only
in Case 5. The Doppler factor $\delta$ exhibits the lowest coverage among all parameters, being below 0.65 in all
cases. These results indicate that flux-based modeling is insufficient for reliably constraining
the Doppler boost for an SSC model for HSPs. External constraint based on the variability time compared
to the light crossing time would remove this parameter and maybe allow for better coverage of other parameters.

The cases with the highest number of parameters (5 out of 7 parameters) with coverage $\geq0.68$ are Cases 3, 7, 9,
14, 15, and 20. Cases 3, 9, and 15 show the strongest high-confidence estimates with 4 parameters with coverage
$\geq0.90$. Interestingly, in Case 3 where only Swift-UVOT, Swift-XRT, and MAGIC data are available $p$,
$\log \gamma_{\rm cut}$, $\log L_e$ and $\log B$ are confidently retrieved. In contrast, Cases 2 shows a poor
coverage for most parameters and only $p$ is recovered above the 0.68 threshold.

To illustrate representative cases, the distributions of the fitted parameters for Cases 2, 3, and 13 are shown
in Figure \ref{fig:kde} (three bottom panels). In Case 3, the distributions of the fitted parameters are narrow
and centered around the true values (red dashed lines), consistent with the high coverage across nearly all
parameters. Case 2, with limited multi-wavelength data, results in broader, less peaked distributions and reduced
reliability for parameter estimates. Case 13 provides an intermediate scenario, where some parameters (e.g., $\log \gamma_{\rm cut}$) remain well constrained while others do not.

\section{Discussion } \label{sec:dis}

Modeling the multiwavelength SEDs of blazars is one of the most effective approaches for understanding the physical properties
of these sources, such as the particle energy distribution, magnetic field strength, Doppler boost and emission region
size. However, the availability of observational data across the full electromagnetic spectrum is often incomplete,
with some energy bands missing due to observational constraints or temporal gaps. Alternatively, understanding which
minimal set of observation bands are required for reliably inferring the largest number of parameters is also important for
multi-wavelength followup strategy. These raise an important question regarding which model parameters can still be reliably
estimated when spectral coverage is incomplete. To address this, we selected three representative sources—one from each
BL Lac blazar subclass (LSP, ISP, HSP)—and modeled their broadband emission within the SSC model. Based on the best-fit models,
we generated synthetic SEDs by systematically simulating different observational configurations that reflect various degrees
of spectral coverage. For each configuration, we produced 1,000 synthetic spectra for each of the 21 observation band
configuration, yielding a total of 63,000 SEDs across the three source types. These datasets, corresponding
to different wavelength availability, were fitted using a CNN surrogate model trained on radiative simulations, enabling
efficient parameter estimation. Compared to the analysis presented in \citet{AOF25}, where the authors studied how different
methods to perform the fit impact the values of the retrieved model parameters of an SSC model, here we systematically
study how the presence and absence of each observation band effect the reliability of parameter recovery.

The fit of the extensive synthetic dataset enables a systematic assessment of the robustness of parameter
recovery based on observational coverage. Namely, this allows to quantify the role of individual spectral bands in
constraining specific physical parameters. By analyzing the impact of data availability on the accuracy and
reliability of recovering model parameters, it is possible to identify for each parameter which observational domains
are critical for effective modeling and which ones contribute less significantly. In the following, we focus on
ISP and HSP blazars, for which the interplay between synchrotron and inverse Compton components occurs across separated
energy bands, making the identification of key constraints more nuanced. The LSP class is excluded from this
discussion, as their SEDs show narrower separation between the synchrotron and inverse Compton peaks and any
combination of optical/UV and X-ray with HE \gray\ data is generally sufficient to recover the model parameters.

\subsection{Bands critical for ISP parameter estimation}

The analysis of ISP blazars shows that reliable parameter recovery is achievable even in configurations with limited
spectral coverage. For the cases with minimal data (Cases 1 to 4), which include only three observational bands, it
is interesting that five out of the seven model parameters exceed the 0.68 coverage threshold  for Cases 1 and 4. This
demonstrates that, despite significant gaps in spectral data, the majority of physical parameters can still be robustly
estimated. In the intermediate cases with four-band coverage (Cases 5 to 8), parameter recovery improves further.
Specifically, Case 7 yields reliable estimates for all parameters, with only the Doppler factor ($\delta$) and the
spectral index $p$ being characterized by a coverage factor smaller than 0.7.

As expected, configurations with broader spectral coverage tend to yield higher coverage values across multiple
parameters. However, this does not imply that simply increasing the number of observed bands will always improve
parameter recovery. In some cases, tighter constraints on one parameter can introduce degeneracies or limit the 
ability to constrain others, highlighting the complex interplay between model parameters. Based on Cases 1 and 4,
the minimal set of energy bands required for effective modeling of ISP blazars includes:
\begin{itemize}
\item[\textit{i)}] a combination of data in the X-ray, HE \gray, and VHE \gray\ bands, in which all parameters can be reliably
recovered except for $p$ and $\log \gamma_{\rm cut}$; and
\item[\textit{ii)}] a combination of data in the optical/UV, X-ray, and HE \gray\ bands, in which all parameters are recoverable
except for $\delta$ and $\log \gamma_{\rm min}$.
\end{itemize}
The inclusion of these bands ensures coverage of both emission components and allows reliable inference of the
underlying physical parameters.

\subsection{Bands critical for HSP parameter estimation}

Modeling the broadband SEDs of HSP blazars generally requires more stringent spectral coverage compared to ISPs.
This is due to the fact that both the synchrotron and inverse Compton components peak at higher energies and parameter
recovery becomes highly sensitive to the presence of data above the X-ray band. Among the configurations with minimal
data coverage (i.e., 3-band observations; cases 1–4), only Case 3 yields reliable estimates (coverage $\geq 0.68$)
for 5 out of 7 model parameters. Specifically, Case 3 includes Swift-UVOT, Swift-XRT and MAGIC. The combined data
sets enable coverage of both the rising and falling segments of the synchrotron component or of the inverse Compton
component in cases that include MAGIC data, thereby allowing a reliable recovery of the majority of the physical
model parameters. The only parameters that are poorly recovered in this case are the Doppler factor ($\delta$) and
the low-energy cutoff of the electron distribution ($\log \gamma_{\rm min}$). This suggests that even minimal
configurations, if strategically composed, can capture the dominant SED features, and allow to confidently recover
several model parameters.

Expanding the dataset to include four-band configurations (Cases 5-8), it is found that Cases 5 and 7 achieves
accurate recovery of 5 parameters. Notably, Case 7 is identical to Case 3, but with the addition of WISE infrared
data. However, the inclusion of IR data does not enhance the recovery of model parameters in this case,
as both $\delta$ and $\log \gamma_{\rm min}$ remain poorly constrained. The lack of constraint on $\log \gamma_{\rm min}$
arises from the fact that it sets a low-energy cutoff in the synchrotron spectrum that lies below the IR regime,
thus falling outside the sensitivity range of the observed bands.

\subsection{Importance of the results for CTAO}

The upcoming CTAO promises to revolutionize our understanding of the HE universe in general, and blazars
in particular \citep{CTAbook}. The energy range of CTAO is comparable to that of MAGIC, though it extends to lower energies (from 20~GeV to 300~TeV) and provides higher sensitivity. Therefore, the results obtained
in this work help identify which additional bands should be combined with CTAO observations to enable the
most reliable parameter recovery. We consider six pairs of Cases—(10,14), (11,15), (12,19), (13,17), (16-20),
and (18,21) —where, in each pair, one case includes MAGIC data and the other does not. The pairs differ in
the additional observational bands included, allowing us to assess the impact of MAGIC data, and by extension
CTAO data in various multi-wavelength configurations.

In the case of ISP blazars, the inclusion of VHE $\gamma$-ray data does not necessarily improve the recovery
of model parameters and, in some configurations, systematically degrades it. This can be attributed to the
spectral position of the VHE $\gamma$-ray band relative to the characteristic SED of ISPs. The inverse Compton
component in these sources typically peaks at energies below the VHE $\gamma$-ray range, meaning that observations
in this band often probe only the steeply falling tail of the HE emission. As a result, data in this band provide
limited constraining power on the peak location or spectral shape, and can even introduce biases when the fit
attempts to reproduce fluxes that are not representative of the dominant emission regime. Moreover, when spectral
coverage already includes data in the infrared, optical/UV, X-ray, and MeV/GeV $\gamma$-ray bands, the additional
information from the VHE $\gamma$-ray band has marginal impact on the overall fit quality. This limitation does
not apply in cases where MeV/GeV data are missing but there is good coverage in the other bands. For example,
in Case 7, where \fermi\ data are absent, the inclusion of VHE $\gamma$-ray observations led to the strongest
parameter recovery. Therefore, for ISPs, the use of VHE $\gamma$-ray data (when MeV/GeV data are available) should
be evaluated with care, as a well-balanced multi-wavelength configuration is often more effective for robust
parameter estimation.

Conversely, in the case of HSP blazars (see Figure \ref{fig:colormap_HSP}), the impact of MAGIC data is evident
through consistent improvements across all pairs considered. For example, comparing Case 10 and Case 14, the
addition of MAGIC data leads to improved estimates of $\log B$, $\log \gamma_{\rm cut}$, $\log R$, $p$ and
$\log L_e$. This indicates that MAGIC data help to better constrain the HE peak of the SED. A more complex
example is the comparison between Case 11 and Case 15, where the inclusion of MAGIC is not required to recover
$\log \gamma_{\rm cut}$, which already has a high coverage probability in Case 11. Interestingly, when broadband
coverage is already extensive—as in Case 12 vs Case 19 and Case 18 vs 21—the impact of MAGIC is reduced. This
suggests that low-energy data tightly constrain the synchrotron component and help define both ends of the
electron energy distribution, making the additional information from MAGIC somewhat redundant in fully sampled
SEDs. As in the ISP case, the Doppler factor $\delta$ and injection Lorentz factor $\log \gamma_{\rm min}$
remain poorly constrained regardless of the inclusion of MAGIC data.

In the modeling, it is important to discuss how individual data in the HE band contribute to constraining model
parameters. We therefore examine whether the MAGIC band alone can provide sufficient constraints in the absence
of \fermi\ data. To investigate this issue, we focus on two pairs of configurations: Cases 3 vs. 4 and Cases 7
vs. 8. In each pair, the only difference is the substitution of data from \fermi\ with MAGIC, allowing for
a direct assessment of the constraining power of MAGIC data when \fermi\ data are unavailable. The pairs differ
by the presence or absence of WISE infrared data, with Cases 3–4 excluding it and Cases 7–8 including it. For the ISP sources,
the results indicate that replacing \fermi\ data with MAGIC data leads to a worse recovery of several key
parameters. Specifically, comparing Cases 3 and 4, the spectral index $p$, magnetic field strength $\log B$,
and electron cutoff energy $\log \gamma_{\rm cut}$ are better constrained in the \fermi\ data -only configuration.
For instance, the recovery probability of $p$ drops from 0.86 (case 4 with \fermi\ data) to 0.53 (case 3 with
MAGIC data), with similar trends for $\log B$ (from 0.85 to 0.68) and $\log \gamma_{\rm cut}$ (from 0.99 to 0.85).
Instead, when WISE data are included (Cases 7 vs. 8), the performance of including MAGIC data are equivalent to
those of \fermi\. In this case, the low energy information provided by WISE help constraining the synchrotron
component at low energy \citep[e.g., see ][]{2024ApJ...963...48G}, and therefore improve the impact of MAGIC
data in the recovery of the parameters. Only the recovery probability of $\log \gamma_{\rm cut}$ worsen, from
0.97 to 0.82. This is because MAGIC data are at higher energy (from 80 GeV to 5 TeV)
than \fermi\ (from 100 MeV to 400 GeV) and does not allow to capture the
peak of the inverse Compton component. Clearly in this case, the improved sensitivity of CTAO at lower energy compared to that of
MAGIC, will enable a better characterization of the inverse Compton scattering component
and improve the parameter recovery performance.

For the HSP sources, the situation is reversed. Replacing \fermi\ data with MAGIC data results in systematically
better recovery of key model parameters. When comparing Case 3 (MAGIC data-only) to Case 4 (\fermi\ data-only),
we observe consistent improvements in coverage across nearly all parameters. Specifically, the recovery probability
of the magnetic field strength $\log B$ increases from 0.58 to 0.92, and the electron luminosity $\log L_e$ also
shows a notable improvement, from 0.65 to 0.95. A smaller improvement is also observed for $\log R$ from 0.57 to
0.82. These differences highlight the critical role MAGIC observations plays in constraining both the synchrotron
and inverse Compton peaks for HSPs. A similar trend is observed when WISE data are included (Cases 7 vs. 8): the
addition of MAGIC data (Case 7) continues to yield better constraints than the \fermi\ data-only configuration
(Case 8). For instance, the recovery probability of $\log B$ improves from 0.63 and 0.9, and that of
$\log \gamma_{\rm cut}$ from 0.80 to 0.98. Overall, for HSP blazars, the presence of MAGIC data proves more effective
than \fermi\ in enabling reliable parameter inference, likely due to the crucial spectral coverage of the
inverse Compton peak provided by MAGIC for this blazar population.

These results indicate a clear divergence in the role of MAGIC observations for ISP and HSP blazars. For ISPs, MAGIC data
alone cannot confidently constrain most SSC model parameters, unless additional constraints are available from
the observations at lower energy bands. In contrast, for HSPs, MAGIC data leads to a systematic improvement in parameter
recovery. This has important implications for future observations with CTAO. The extended sensitivity of CTAO at lower
energies (compared to MAGIC), reaching down to tens of GeV and partially overlapping with the sensitivity window of \fermi,
will be particularly important for ISP observations. It will enable robust parameter inference even when \fermi\ data
are unavailable. For HSPs, CTAO is expected to provide comparable or improved performance relative to MAGIC, supporting
accurate modeling of blazar high-energy emission.

\section{Conclusions} \label{sec:conc}

In this work, we studied how the availability of multiwavelength observational data affects the reliability
of parameter recovery within the SSC framework for three representative blazars: OJ 287 (LSP), TXS 0506+056 (ISP), and Mrk
421 (HSP). Using synthetic datasets generated under different spectral coverage scenarios, we quantified the degree to which
physical model parameters can be recovered from flux-based SED fitting.

We have shown that for LSP blazars, a minimal combination of Swift-UVOT, -XRT, and \fermi\ data is sufficient to robustly
constrain most SSC model parameters. For ISP blazars, a minimal set of observational bands capable of yielding reliable
estimates for a significant number of parameters (5 out of 7) consists of either X-ray, HE $\gamma$-ray, and VHE $\gamma$-ray data,
or optical/UV, X-ray, and HE $\gamma$-ray data. Similarly, for HSPs, the combination of optical/UV, X-ray, and VHE
$\gamma$-ray data is sufficient to recover the majority of model parameters reliably. The contribution of VHE $\gamma$-ray
data to parameter recovery differs between ISPs and HSPs. For ISPs, VHE $\gamma$-ray observations, when combined with
lower-energy data (e.g., infrared), can effectively constrain model parameters even in the absence of HE $\gamma$-ray data.
In contrast, for HSPs, VHE $\gamma$-ray data often encompass the peak of the high-energy component, leading to significant
improvements in parameter recovery even when HE $\gamma$-ray data are unavailable.

To conclude, this study provides a practical road-map for the design of future multi-wavelength observation campaigns,
identifying the minimal set of instruments needed to reliably constrain key physical parameters for each blazar subclass.
It also serves as a diagnostic framework when modeling specific observational periods, allowing researchers to assess
which parameters can be reliably inferred based on the available data.

\begin{acknowledgments}
NS acknowledges the support by the Higher Education and Science
Committee of the Republic of Armenia, in the frames of the
research project No 23LCG-1C004.
\end{acknowledgments}










\bibliography{biblio}{}
\bibliographystyle{mnras}



\end{document}